\documentclass[12pt]{article}
\usepackage[dvips]{graphicx}
\begin{document}
\title {Inelastic scattering of protons from {\Large $^{6,8}$}He and 
{\Large $^{7,11}$}Li in a folding model approach}  
\author{
D. Gupta$^1${\footnote{\small Corresponding Author Email: 
dhruba@hp1.saha.ernet.in}} and 
C. Samanta$^{1,2}${\footnote{\small Email: chhanda@hp1.saha.ernet.in}}\\
$^1$Saha Institute of Nuclear Physics,\\ 1/AF, Bidhannagar, 
Calcutta 700064, India\\
$^2$ Physics Department, Virginia Commonwealth University,\\ Richmond,
Virginia 23284-2000, USA}
\maketitle
\begin{abstract}
The proton-inelastic scattering from $^{6,8}$He and $^{7,11}$Li nuclei 
are studied in a folding model approach. A finite-range, momentum, density and 
isospin dependent nucleon-nucleon interaction (SBM) is folded with realistic 
density distributions of the above nuclei. The renormalization factors N$_R$ 
and N$_I$ on the real and volume imaginary part of the folded potentials are 
obtained by analyzing the respective elastic scattering data and kept unaltered 
for the inelastic analysis at the same energy. The form factors are generated 
by taking derivatives of the folded potentials and therefore required 
renormalizations. The $\beta$ values are extracted by fitting the p + 
$^{6,8}$He,$^{7,11}$Li inelastic angular distributions. The present analysis 
of p + $^8$He inelastic scattering to the 3.57 MeV excited state, including 
unpublished forward angle data (RIKEN) confirms L = 2 transition. Similar 
analysis of the p + $^6$He inelastic scattering angular distribution leading 
to the 1.8 MeV (L = 2) excited state fails to satisfactorily reproduce the data.
\vskip .2 true cm
PACS number(s): 25.40.Cm, 25.40.Ep, 25.60.-t, 21.45.+v, 21.10.Gv   
\vskip .3 true cm
\end{abstract}
\newpage
{\bf 1. Introduction}
\vskip .2 true cm
Recent advances in nuclear physics have given us the opportunity to delve into 
unique problems, hitherto unknown. An example is the neutron halo in the
nucleus $^{11}$Li, discovered as a consequence of its very large interaction
radius, deduced from the measured interaction cross sections of $^{11}$Li
with various target nuclei~\cite{TA85,TA88,MI87}. The halo of the nucleus 
extends its matter distribution to a large radius. Thus, the two valence 
neutrons in $^{11}$Li, which form the halo, are extended well beyond the $^9$Li 
core and the 2n separation energy is exceedingly small (0.247 MeV). Besides the 
renowned example of $^{11}$Li, there are other neutron rich nuclei, like $^6$He 
and $^8$He having extended valence neutron distributions called neutron
halos/skins~\cite{TA92,GO95}. While $^6$He and $^{11}$Li have predominant
core and two valence neutron structures, the $^8$He nucleus has four valence 
neutrons and has the largest neutron to proton ratio among these three nuclei.

Along with the structures of such exotic nuclei near the drip lines, their 
excitation modes have also attracted considerable attention in recent 
times~\cite{GO95,KO972,NA00}. The existence of the neutron halo around the 
core, gave birth to the idea of a new resonance mode of excitation called, the 
soft dipole resonance (SDR)~\cite{HA87,IK92}, in which the halo neutrons 
oscillate against the core nucleus. From phenomenological and microscopic
analyses of the $^{11}$Li(p,p')$^{11}$Li$^*_{1.3}$ angular distribution data,
excitation of the 1.3 MeV resonant state of $^{11}$Li was found to correspond 
to L = 1 transition~\cite{KO972,KA97}, arising from SDR~\cite{KA98,KA99}.
Earlier microscopic analysis of p + $^7$Li inelastic scattering, include the 
work of Mani et al.~\cite{MA712} and Petrovich et al.~\cite{PE93}, where 
abnormally large deformations were predicted in~\cite{MA712}. For p + $^8$He 
inelastic scattering, coupled channel calculations using a Woods Saxon 
potential was carried out in Ref.~\cite{GO95} and spin-parity of the excited 
state (E$^*$ = 3.57 MeV) was predicted. The  p + $^6$He inelastic angular 
distribution data leading to E$^*$ = 1.8 MeV~\cite{AUprivate} is also 
available. Since all of $^{7,11}$Li and $^{6,8}$He are loosely bound, their 
wave functions are quite extended in space. But, their difference in internal 
structures could lead to different excitation modes. In fact, SDR has been 
predicted in $^6$He and $^{11}$Li, but not in $^7$Li and $^8$He. To appreciate 
this difference, a systematic study of their inelastic proton scattering data 
is desirable. 

In this work we present a consistent analysis of the proton inelastic 
scattering from $^{7,11}$Li and $^{6,8}$He nuclei in a folding model approach.
Although sophisticated folding calculations of two nucleon t- and g-matrices 
exist in a recent review of Amos et al.~\cite{AM00}, we feel that, a simpler 
calculation is always useful. The folding model is well known as a powerful 
tool for analyzing nucleus-nucleus scattering data at relatively low incident 
energies~\cite{PE93,RE74,SA97,GU002}. It directly links the density profile 
of the nucleus with the scattering cross sections and is thus very appropriate 
for studying nuclei with exotic matter distributions. However, in such analysis 
the choice of the nucleon-nucleon interaction is very crucial. As the unstable 
radioactive nuclei have different neutron and proton density distributions, 
an isospin sensitive nucleon-nucleon interaction is required to construct the 
folded potentials~\cite{KA97,CH94}. For nuclei with low breakup thresholds, 
the folding model analysis also gives an estimation of the breakup channel 
coupling effects~\cite{SA83} on the elastic through its requirement of a 
renormalization factor (N$_R$)~\cite{SA79} on folded potentials. With 
increasing incident energy, the breakup channel coupling effect decreases, 
and the value of N$_R$ approaches 1.0.

Using a density, momentum and isospin dependent finite range effective
interaction~\cite{BA90} in a single folding model, the present work analyses
the available low energy  proton inelastic scattering data from $^7$Li 
(49.75A MeV)~\cite{MA712}, $^{11}$Li (68.4A MeV)~\cite{KO972}, $^8$He 
(72.5A MeV)~\cite{KO93,KOprivate} and $^6$He (40.9A MeV)~\cite{AUprivate}.
A semi-microscopic analysis in the optical model (OM) framework is carried out 
for the p + $^7$Li,$^6$He elastic scattering data while the OM analysis of p + 
$^{11}$Li,$^8$He elastic scattering at the above energies has already 
been performed on the same footing in the earlier work~\cite{GU002}. In the 
DWBA calculations of the nuclear excitation, with transferred angular 
momentum L, the form factors used are obtained by taking the derivative of 
the semi-microscopic potentials used. The p + $^{11}$Li inelastic scattering 
though analyzed in~\cite{KA97,KA98,KA99} is repeated here with microscopic 
real and volume imaginary potentials in addition to phenomenological surface 
and spin-orbit potentials, and by generating conventional form factors. 
Slightly different N$_R$ and N$_I$ values were obtained by a $\chi^2$ fit 
in~\cite{GU002} and these new values are used in the present work for the 
sake of completeness and comparison of the results with other nuclei. 
The formalism and the analysis are given in section 2 while the summary and 
conclusions are given in section 3.

\vskip 1. true cm
{\bf 2. Formalism and Analysis}
\vskip .3 true cm

The form of the single folded potential~\cite{SA79}, used in the present work 
is, \\${\rm U(\vec{r_1}) = 
\int \rho(\vec{r_2}) v_{NN}(|\vec{r_1}-\vec{r_2}|)d^3\vec{r_2}}$\\
where, $\rho(\vec{r_2})$ is density of the nucleus and v$_{\rm NN}$ is the 
effective interaction between two nucleons at the sites $\vec{r_1}$ and 
$\vec{r_2}$ with densities $\rho_1(\vec{r_1})$ and $\rho_2(\vec{r_2})$ 
respectively. A finite-range, density, momentum and isospin dependent 
effective interaction SBM (Modified Seyler Blanchard) is chosen, which has 
different strengths for pp (or nn) and pn interactions and its form 
is~\cite{BA90},\\
${\rm v_{eff} (r = |\vec{r_1} - \vec{r_2}|,p,\rho) 
= -C_{l,u}\frac{e^{-r/a}}{r/a}[1 - \frac{p^2}{b^2} -d^2(\rho_1 +\rho_2)^n]}$,\\
where, the subscripts `l' and `u' refer to like-pair (nn or pp) and unlike-pair 
(np) interactions, respectively. Here `a' is the range of the two-body 
interaction in the configuration space, `b' is a measure of the strength of 
repulsion with relative momentum `p', while `d' and `n' are two parameters 
determining the strength of density dependence. The parameters n, C$_{\rm l}$, 
C$_{\rm u}$, a, b, d are given in Table 1. These constants are found to 
reproduce the bulk properties of nuclear matter and of finite 
nuclei~\cite{BA90,SA89} and are known also to explain the p + 
$^{4,6,8}$He,$^{6,7,9,11}$Li scattering data 
successfully~\cite{KA97,KA98,KA99,GU002,KA95}. The parameters are determined
without exchange effects and thus they contain the effect indirectly though
in a very approximate way.

The $^{6,8}$He and $^{7,11}$Li densities used in this work are shown in Fig. 1. 
The density prescriptions remain the same as that used in the earlier work on 
elastic proton scattering from these nuclei~\cite{GU002}. For $^{11}$Li, the 
cluster orbital shell model (COSM) density~\cite{KA97,OG92} has been used. The 
parametric form of the $^7$Li density is used from Ref.~\cite{PE93}. The 
density of $^8$He~\cite{KO971}, was also derived in the COSM approximation. 
It contains the extended distribution of valence nucleons and correspond to 
the experimental matter radius. For $^6$He, the p-inelastic scattering data 
at 40.9A MeV is recently available~\cite{AUprivate} and the L-transfer values 
are already predicted for some excited states~\cite{NA00,AJ88}. In this work, 
a number of ground state densities derived~\cite{AL96,THprivate,ZH93} by using 
Faddeev wave function models called, P1, FC, FC6, Q3, Q1, FB, FA, K ,C, are 
employed to predict angular distributions for excitation to the 1.8 MeV 
state. Those density models incorporate different n-n and n-$\alpha$ 
potentials with a variation of the two-neutron separation energy E(2n) from 
about -1.15 MeV to -0.21 MeV and thereby a variation of the root mean squared 
(rms) radius of $^6$He. The rms radii corresponding to the above models are 
2.32, 2.50, 2.53, 2.54, 2.56, 2.64, 2.64, 2.66, 2.76 fm respectively. These 
radii were computed assuming that the bare $^4$He core rms radius is 1.49 
fm~\cite{THprivate}. The $^4$He density is also plotted in Fig. 1 to show that 
all the nuclear densities have a tail extended well beyond the $\alpha$-core. 
Since the interaction is isospin sensitive, separate neutron and proton 
densities of the nuclei are used~\cite{GU002,KO971,GU001} for folding 
calculations.

Both the real (V) and volume imaginary (W) parts of the potentials (generated
microscopically by folding model) are assumed to have the same shape, as in 
Ref.~\cite{GU002}, i.e. V$_{micro}$(r) = V + iW = 
(N$_R$ + iN$_I$)U(r$_1$) where, N$_R$ and N$_I$ are the renormalization factors 
for the real and imaginary parts respectively~\cite{SA97}. These folded 
potentials with appropriate N$_R$ and N$_I$ as required for elastic scattering 
fits (Table 3 of Ref.~\cite{GU002} and this work), are used subsequently for 
inelastic scattering analysis in this work. The spin-orbit and the surface 
imaginary parts are taken from the phenomenological best fit calculations as 
before (Table 2 of Ref.~\cite{GU002} and this work). They needed minor 
adjustments in some cases for best fits as reported earlier~\cite{GU002} and 
in this work. The phenomenological potentials had the following
form,\\ V$_{pheno}$(r) = -V$_o$~f$_o$(r)~-~i~W$_v~$f$_v$(r) + 
4~i~a$_s$W$_s$(d/dr) f$_s$(r) + 2($\hbar$/m$_{\pi}$c)$^2~V_{s.o}$ 1/r 
(d/dr) f$_{s.o}$(r)~({\bf L.S}) + V$_{coul}$,\\
where, f$_{\rm x}$(r)~= [1~+~exp($\frac{\rm r-R_x}{\rm a_x}$)]$^{-1}$ 
and R$_{\rm x}$~=~r$_{\rm x}$A$^{1/3}$. The subscripts $o,v,s,s.o$ denote
real, volume imaginary, surface imaginary and spin-orbit respectively and 
V$_o$, W$_v$ (W$_s$) and V$_{s.o}$ are the strengths of the real, volume 
(surface) imaginary and spin-orbit potentials respectively. V$_{coul}$ is the 
Coulomb potential of a uniformly charged sphere of radius 1.40~A$^{1/3}$.
For each angular distribution studied before~\cite{GU002} as well as here, 
best fits are obtained by minimizing $\chi^2$/N, where $\chi^2$ = 
$\sum_{k = 1}^{\rm N} \left[\frac{\sigma_{th}(\theta_k)~-
\sigma_{ex}(\theta_k)}{\Delta\sigma_{ex} (\theta_k)}\right]^2$, $\sigma_{th}$/
$\sigma_{ex}$ are the theoretical/experimental cross sections at angle 
$\theta_k$, $\Delta\sigma_{ex}$ is the experimental error and N is the number 
of data points. 

The best fit OM parameters for the p + $^7$Li elastic scattering at E = 49.75A 
MeV are given in Table 2. In the present semi-microscopic analysis, a search 
on N$_R$ and N$_I$ is carried out for minimum $\chi^2$/N and the values are 
given in Table 3 and the fit is shown in Fig. 2a. The surface imaginary and 
spin-orbit potentials remain the same as obtained from the phenomenological 
best fits. For the p + $^8$He,$^{11}$Li elastic scattering data at 72.5A and 
68.4A MeV respectively, the required N$_R$, N$_I$ and phenomenological 
potentials (W$_s$ and V$_{s.o}$) are already available from the earlier 
analysis~\cite{GU002}. Using OM parameter setII of p + $^8$He elastic 
scattering~\cite{GO95} yields much lesser N$_R$ value and a better fit to the 
inelastic data. The corresponding best fit parameters for p + $^6$He scattering
at 40.9A MeV are also given in Table 2,3. To yield minimum $\chi^2$/N, the 
r$_s$ value had to be changed from 1.6 to 1.43, 1.32 and 1.26 fm for the P1, Q1
and C density models of $^6$He. These potentials are therefore used in the DWBA 
calculations of inelastic scattering with transferred angular momentum L. The 
calculations are performed by using the code DWUCK4~\cite{DWUCK4}. The 
conventional form factors, i.e. derivative of the potentials are used. The 
microscopic real and imaginary form factors have the same shape with strengths 
N$_R^{FF}$ and N$_I^{FF}$ respectively, where N$_{R,I}^{FF}$ = 
N$_{R,I}r_{rms}^V$, where the radius parameter $r_{rms}^V$ is the rms 
radius of the folded potential. Thus the renormalization of the form factors 
is consistent with that for the folded potential. Form factors derived from 
phenomenological surface imaginary and spin-orbit potentials are also included.

To fit the p + $^7$Li inelastic angular distributions leading to the 0.478 
and 4.63 MeV excited states of $^7$Li, N$_{R,I}$ values from elastic 
scattering fits and the corresponding N$_{R,I}^{FF}$ are employed. In the 
former, the best fit yields for the deformation parameter $\beta$ a value of 
0.59 (Fig. 2b, Table 3) for transferred angular momentum L = 2 
(3/2$^-$ to 1/2$^-$). For the 4.63 MeV excited state of $^7$Li best fit gives 
a $\beta$ = 0.82 for L = 2 (3/2$^-$ to 7/2$^-$). The calculations could 
explain the data up to $\theta_{cm}$ $\sim$ 110$^o$ and $\chi^2$ value is 
calculated by incorporating the data points only up to that (Fig. 2c, Table 3). 
Earlier works~\cite{KO972,KA97,KA98,KA99} showed that the p + $^{11}$Li 
inelastic scattering data at E = 68.4A MeV could be satisfactorily explained
for angular momentum transfer L = 1. The N$_R$ = 0.50 and N$_I$ = 0.15 values 
are used in the present work as obtained from earlier work on p-elastic 
scattering~\cite{GU002}. The $\chi^2$ minimum test for best fit resulted in 
a $\beta$ value of 0.58 (Fig. 3, Table 3), with L = 1. Changing the form 
factors generated from the surface imaginary and spin-orbit phenomenological 
potentials have negligible effects on the angular distribution. 

The $^8$He nucleus has a very high neutron-to-proton ratio. To confirm the 
spin-parity of the excited state E$^*$ = 3.57 MeV, studies over wide angular 
range are carried out. We try to fit both the reported $^8$He$^*$ data 
(setI)~\cite{KO93} as well as the unpublished angular distribution data 
(setII)~\cite{KOprivate}. The setII data were obtained from invariant mass 
measurements as reported in~\cite{KO93}. They were measured not by proton 
detection (like above), but by detecting $^6$He + n + n coincidences. Namely, 
the shape of the angular distribution for the inelastic scattering was 
extracted, while absolute value of the cross section was simply normalized in 
consistency with the above given inelastic data from proton measurements. It 
is seen that L = 2 (J$^\pi$ = 2$^+$) gives best fit. L = 1 is excluded by the 
new measurement (setII) and L = 3 is excluded by all experimental points at 
larger angles (Fig. 4, Table 3). The $\beta$ = 0.28 is lesser than 0.44 
in~\cite{GO95}.

Similar analysis is extended to the recently available p + $^6$He inelastic 
angular distributions at 40.9A MeV (Fig. 5). The Faddeev wave function 
densities of $^6$He~\cite{AL96,THprivate} are employed and the N$_R$, N$_I$ 
values extracted from elastic scattering (Fig. 5a) are used. The angular 
distribution for the 1.8 MeV (J$^\pi$ = 2$^+$) excitation is shown in Fig. 5b.
For the sake of clarity only three calculations (P1, Q1, C models) 
are plotted and the rest 
lie between P1 and C. It is found that contrary to very good fits to the 
inelastic data for p + $^8$He,$^{7,11}$Li, the present formalism is unable to 
give satisfactory fit to the p + $^6$He data. But the $\beta$ value extracted 
from the optimized fit agrees closely with that of Aumann et al.~\cite{AU99} 
and Rusek et al.~\cite{RU00} ($\beta$ $\sim$ 0.78). It is observed that though 
the three $^6$He densities correspond to a variation of rms radius from 2.32 
fm to 2.76 fm, the corresponding change in the angular distributions of 
inelastic scattering is negligible.

\vskip 1. true cm
{\bf 3. Summary and Conclusions}
\vskip .3 true cm

A consistent folding model analysis of p-inelastic scattering of stable and
unstable nuclei can provide valuable insight into their structure and reaction 
dynamics. The present work aims at studying the inelastic scattering of protons 
from $^{6,8}$He and $^{7,11}$Li nuclei. The semi-microscopic analysis followed 
here, involves a finite-range, momentum, density and isospin dependent 
nucleon-nucleon effective interaction (SBM) and realistic densities of 
different nuclei. Earlier, a similar analysis of p-elastic scattering data on 
these nuclei showed that renormalizations (N$_R$ and N$_I$)  of the real and 
volume imaginary part of the folded potentials~\cite{GU002} are required.
These factors, once determined from the elastic data, are kept unaltered for 
the inelastic data analysis to ensure a model independent study.

The conventional way of generating the form factors is followed, i.e., by 
taking the derivatives of the potentials (microscopic real and imaginary as 
well as phenomenological surface imaginary and spin orbit). Deformation 
parameters ($\beta$) are extracted from the analyses. The 
unpublished~\cite{KOprivate} forward angle data of p + $^8$He inelastic 
angular distribution shows that best fit implies an L = 2 transition 
(i.e J$^\pi_{3.57}$ = 2$^+$) whereas L = 1, 3 fail to reproduce the exact 
structure. 

Analysis on the same footing enables us to study elastic and inelastic
p + $^6$He scattering at E = 40.9A MeV. In contrast to the other nuclei the 
p + $^6$He inelastic scattering to the 1.8 MeV state
(J$^\pi_{1.8}$ = 2$^+$, L = 2) could not be satisfactorily explained by the
present formalism, though the extracted $\beta$ value agrees closely
with previous works~\cite{AU99,RU00}. Moreover, inclusion of various ground 
state density prescriptions of varying r.m.s radii (from 2.32 fm to 2.76 fm) 
have negligible influence on the inelastic observables studied here.

The authors gratefully acknowledge A. A. Korsheninnikov for sending his 
unpublished experimental data in a tabular form and giving his 
consent to use it in this paper. Thanks to I. J. Thompson for providing the 
$^6$He densities, F. Auger for sending the data in tabular form, R. Kanungo 
for the folding model code and S. K. Samaddar
for useful discussions. D.G. acknowledges the CSIR, India for financial support.

\begin{table}
{\bf Table 1:} \\
Parameters of the SBM interaction in MeV-fm units \\
\\
\begin{tabular}{cccccc} 
\hline
\multicolumn{1}{c}{n}&
\multicolumn{1}{c}{C$_{\rm l}$}&
\multicolumn{1}{c}{C$_{\rm u}$}&
\multicolumn{1}{c}{a}&
\multicolumn{1}{c}{b}&
\multicolumn{1}{c}{d}\\
\hline
2/3&215.7&669.3&0.554&668.7&0.813\\
\hline
\end{tabular}
\end{table}

\newpage
\begin{table}
{\bf Table 2:} \\
Optical potential parameters used in p + nucleus elastic scattering \\
\\
\setlength{\tabcolsep}{0.2 mm}
\begin{tabular}{c|c|c|c|c|c|c|c|c|c|c|c|c|c|c|l}
\hline
\multicolumn{1}{c}{Nucleus}&
\multicolumn{1}{|c|}{E/A}&
\multicolumn{1}{|c}{V$_o$}&
\multicolumn{1}{c}{r$_o$}&
\multicolumn{1}{c|}{a$_o$}&
\multicolumn{1}{c}{W$_v$}&
\multicolumn{1}{c}{r$_v$}&
\multicolumn{1}{c|}{a$_v$}&
\multicolumn{1}{c}{W$_s$}&
\multicolumn{1}{c}{r$_s$}&
\multicolumn{1}{c|}{a$_s$}&
\multicolumn{1}{c}{V$_{s.o}$}&
\multicolumn{1}{c}{r$_{s.o}$}&
\multicolumn{1}{c|}{a$_{s.o}$}&
\multicolumn{1}{c|}{J/A}&
\multicolumn{1}{l}{ Ref.}\\
&(MeV)&(MeV)&(fm)&(fm)&(MeV)&(fm)&(fm)&(MeV)&(fm)&(fm)&(MeV)&(fm)&(fm)&(MeV fm$^3$)&\\
\hline
$^7$Li&49.8&36.70&1.210&0.550&5.62&1.730&1.220&&&&4.90&1.000&0.530&424.6&\cite{MA711}\\
$^{11}$Li&68.4&14.50&1.385&0.546&4.26&0.560&1.160&4.60&0.560&1.160&5.90&0.800&0.630&211.5&\cite{GU002}\\
$^8$He&72.5&21.60&0.970&0.857&1.43&0.820&0.633&1.50&1.600&0.717&6.39&1.430&0.801&246.1&\cite{GO95}\\
$^6$He&40.9&45.40&0.990&0.612&2.60&1.101&0.690&3.47&1.600&0.772&5.90&0.677&0.630&397.7&[*]\\
\hline
\end{tabular}
[*] this work
\\
\end{table}
 
\begin{table}
{\bf Table 3:} \\
Renormalizations of 
SBM folded potentials and form factors for p-nucleus scattering at 
incident energy (E/A) and excited state energy (E$^*$) in MeV,
angular momentum transfer (L), deformation parameter ($\beta$), volume integral
(J/A) of the real folded potential in MeV fm$^3$ and $\chi^2$/N values from the 
elastic and inelastic scattering best-fits\\

\begin{tabular}{ccccccccccccc} 
\hline
\multicolumn{1}{c}{Nucleus}&
\multicolumn{1}{c}{E/A}&
\multicolumn{1}{c}{E$^*$}&
\multicolumn{1}{c}{N$_R$}&
\multicolumn{1}{c}{N$_I$}&
\multicolumn{1}{c}{r$_{rms}^V$}&
\multicolumn{1}{c}{N$_R^{FF}$}&
\multicolumn{1}{c}{N$_I^{FF}$}&
\multicolumn{1}{r}{L}&
\multicolumn{1}{r}{$\beta$}&
\multicolumn{1}{r}{$\chi^2_{el}$/N}&
\multicolumn{1}{r}{$\chi^2_{inel}$/N}&
\multicolumn{1}{r}{J/A}\\
\hline
$^7$Li$^*$&49.8&0.478&0.75&0.24&2.853&2.140&0.685&2&0.59&5.681&6.692&423.8\\
$^7$Li$^*$&49.8&4.630&0.75&0.24&2.853&2.140&0.685&2&0.82&5.681&1.369&423.8\\
\hline
$^{11}$Li$^*$&68.4&1.300&0.50&0.15&3.909&1.954&0.586&1&0.58&0.488&0.353&282.6\\
\hline
$^8$He$^*$&72.5&3.570&0.41&0.00&3.300&1.353&0.000&1&0.32&0.594&37.007&247.7\\
$^8$He$^*$&72.5&3.570&0.41&0.00&3.300&1.353&0.000&2&0.28&0.594&0.452&247.7\\
$^8$He$^*$&72.5&3.570&0.41&0.00&3.300&1.353&0.000&3&0.55&0.594&24.777&247.7\\
\hline
$^6$He$^*$(P1)&40.9&1.800&0.68&0.00&3.125&2.125&0.000&2&0.71&6.247&11.361&444.1\\
$^6$He$^*$(Q1)&40.9&1.800&0.68&0.00&3.358&2.283&0.000&2&0.71&6.907&14.743&469.8\\
$^6$He$^*$(C)&40.9&1.800&0.68&0.00&3.558&2.419&0.000&2&0.71&7.980&18.178&484.8\\
\hline
\end{tabular}
\end{table}

\newpage
{\bf Figure Captions}

\begin{enumerate}

\item{The densities of (a) $^{4,6,8}$He (b) $^4$He, $^{7,11}$Li
used in this work (see the text for references).}

\item{The experimental angular distributions and folding model calculations
(employing SBM interaction) of 
$^7$Li at 49.75A MeV for (a) elastic and (b) E$^*$ = 0.478 MeV (1/2$^-$), 
(c) E$^*$ = 4.63 MeV (7/2$^-$) state for inelastic scattering. 
Here L = 2 in (b) and (c).
The corresponding N$_R$, N$_I$, N$_R^{FF}$, N$_I^{FF}$ values and
phenomenological surface imaginary and spin-orbit parameters are given 
in Table 2, 3.}

\item{The same as in Fig. 2 for 
$^{11}$Li at 68.4A MeV and E$^*$ = 1.3 MeV state for inelastic scattering. 
Here L = 1.}

\item{The same as in Fig. 2 for 
$^8$He at 72.5A MeV for (a) elastic and (b) 
E$^*$ = 3.57 MeV state for inelastic scattering.
In (b) the data represented by solid circles (Set I) are from~\cite{KO93} 
while the data represented by hollow circles (Set II) are yet unpublished and 
obtained from~\cite{KOprivate}. The 
L = 1 , 2 , 3 calculations are shown by dashed, solid
and dotted curves respectively.} 

\item{The same as in Fig. 2 for 
$^6$He at 40.9A MeV for (a) elastic and (b) 
E$^*$ = 1.8 MeV state for inelastic scattering. 
Here L = 2. The calculations are shown by the dashed,
solid and dotted curves for the P1, Q1 and C models.}

\end{enumerate}

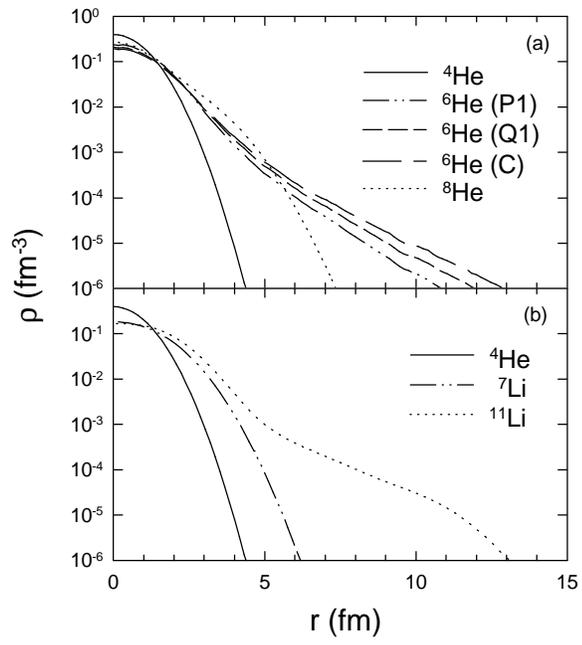
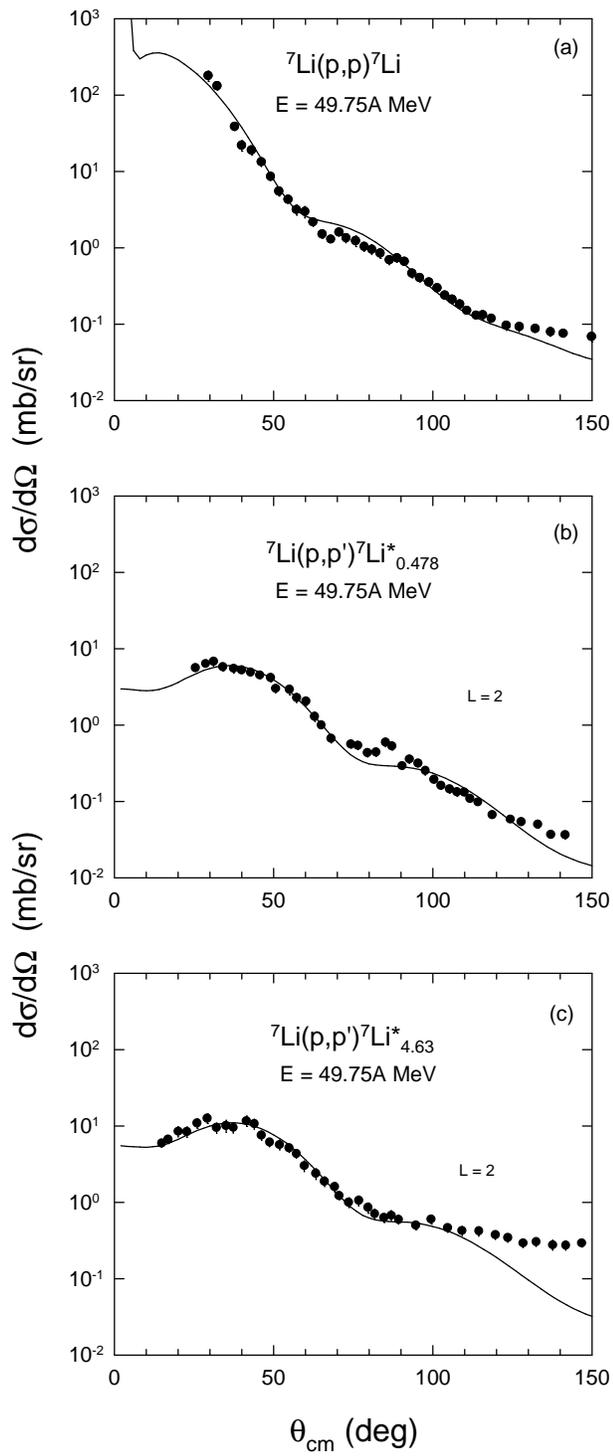
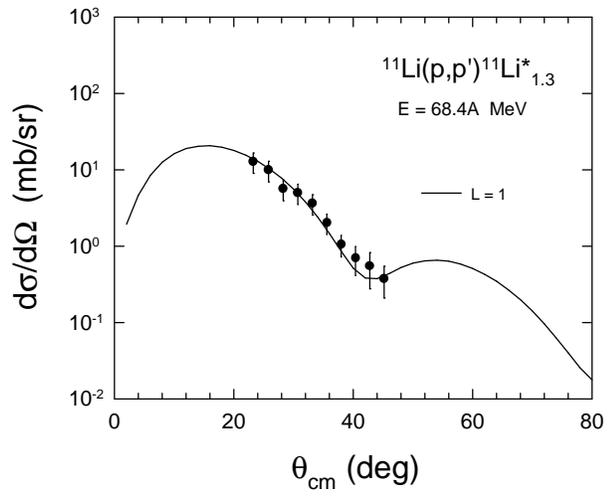
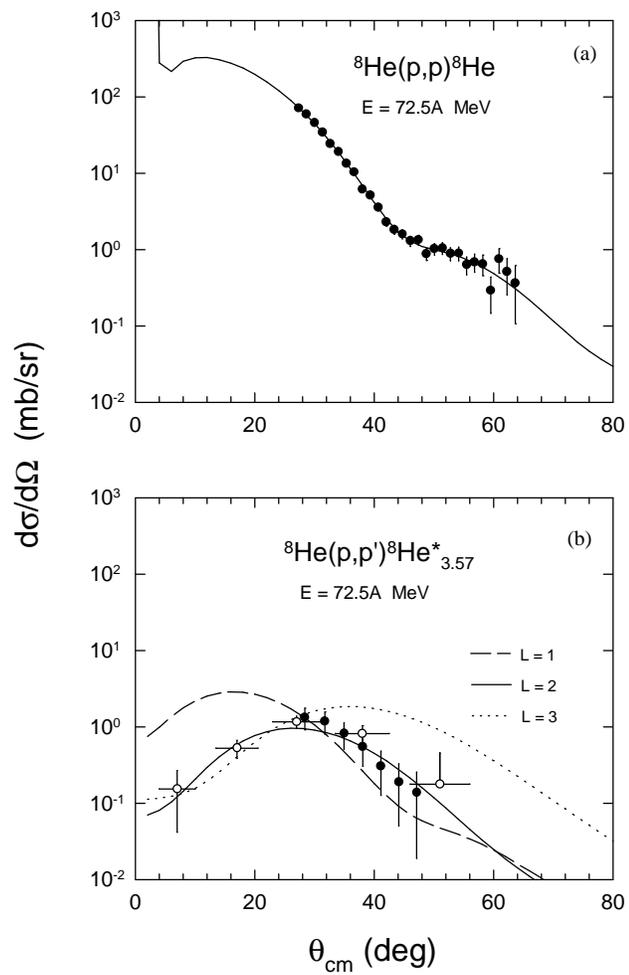
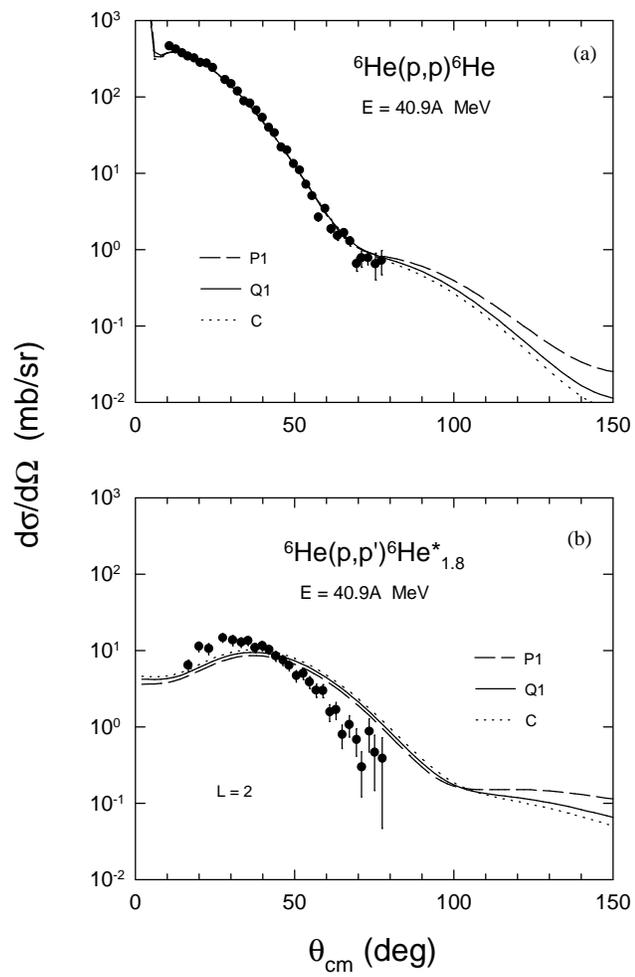
\end{document}